\documentclass{tlp}
\usepackage{latexsym}
\usepackage{amssymb}
\begin{document}
\title[Three Optimisations for Sharing]
{Three Optimisations for Sharing}
\author[Jacob M.~Howe and Andy King]
{JACOB M.~HOWE\\
Department of Computing, City University,\\
London, EC1V OHB, UK. \\
email: \texttt{jacob@soi.city.ac.uk} \and
ANDY KING\\
Computing Laboratory,
University of Kent at Canterbury,\\
Canterbury, CT2~7NF, UK. \\
email: \texttt{a.m.king@ukc.ac.uk}}

\maketitle

\newtheorem{lemma}{Lemma}[section]
\newtheorem{proposition}{Proposition}[section]
\newtheorem{theorem}{Theorem}[section]
\newtheorem{corollary}{Corollary}[section]
\newtheorem{definition}{Definition}[section]
\newtheorem{example}{Example}[section]

\def\Gr     {Gr}
\def\Share  {Sh}
\def\Free   {Fr}
\def\Lin    {Lin}
\def\nat    {\mathbb{N}}
\def\opp    {-}
\def\shr    {\sim}
\def\reduce{\leadsto}

\def\myshare    {\mbox{\it Sharing\/}}
\def\mymodel    {\mbox{\it model\/}}
\def\mybool {\mbox{\it Bool\/}}
\def\mypos  {\mbox{\it Pos\/}}

\begin{abstract}
In order to improve precision and efficiency sharing analysis
should track both freeness and linearity. The abstract unification
algorithms for these combined domains are suboptimal, hence there
is scope for improving precision. This paper proposes three
optimisations for tracing sharing in combination with
freeness and linearity.
A novel connection between equations and sharing
abstractions is used to establish correctness
of these optimisations even in the presence of rational trees.
A method for pruning intermediate
sharing abstractions to improve efficiency is also proposed. The
optimisations are lightweight
and therefore
some, if not all, of these optimisations will be of interest to the
implementor.

\end{abstract}

\section{Introduction}

A set-sharing analyser will usually also track freeness and
linearity. This is because freeness and linearity are cheap to
maintain and result in more accurate, that is smaller, sharing
abstractions which in turn improve the efficiency of the sharing
component of abstract unification. However, current abstract
unification algorithms for sharing, freeness and linearity are
suboptimal. This paper considers how to improve the precision of
sharing with freeness and linearity by considering the interaction
of these components. These refinements do not incur a significant
computational overhead.  To this end three optimisations are
given, along with examples of where precision is gained.  Their
cost is discussed and correctness proved. 

The first optimisation follows from the observation that the
algorithms for pair-sharing with linearity can sometimes
out perform set-sharing with linearity (in terms of which pairs of
variables may share). This is because of an independence check
which pervades the set-sharing literature (from early work
\cite{L91} to the most recent and comprehensive \cite{BZH00}).
This check is in fact redundant. By removing this,
the precision of abstract unification is improved, since
linearity can be exploited more frequently.

Fil{\'e} \cite{F94} observed that freeness can be used to
decompose a sharing abstraction into a set of sharing abstractions.
For each component of the decomposition, the sharing groups of 
that component do not (definitely) arise from different computational
paths. Abstract unification can then be
applied to each component and the resulting abstractions merged.
This tactic has not been included in analysers owing to its
prohibitive cost. The second optimisation is a lightweight
refinement of abstract unification inspired by the decomposition.
Though not as precise as the full decomposition, it does achieve
the necessary balance between cost and benefit.

Thirdly, an optimisation for pruning sharing groups is presented.
This tactic demonstrates that sharing in combination with freeness
can improve groundness which, in turn, can improve sharing (even
in the presence of rational trees). Put another way, it means that
any optimal algorithm for sharing, freeness and linearity will
have to consider subtle interactions between sharing, freeness and
groundness.

One principle of set-sharing is that the number of sharing groups
should be minimised. As well as increasing precision, this can
improve efficiency and possibly avoid widening. A fourth technique
is proposed which can prune the size of inputs to the
abstract unification algorithm by considering the grounding
behaviour of sets of equations. Reducing the size of the inputs
(and intermediate abstractions) simplifies abstract unification
and can thereby improve performance.  Whilst the technique will
not theoretically
improve the precision of the overall result, in practice,
a precision gain might be
achieved if widening is avoided within the unification algorithm.

Correctness is expressed in terms of a
novel concretisation map which characterises equations as their
idempotent most general unifiers. This simplifies the correctness
arguments and in particular enables the
abstract unification algorithms to be proved correct
for rational tree constraint solving (as adopted by SICStus
Prolog and Prolog-III). 
To the best of the authors' knowledge, this is the
first proof of correctness for a sharing, freeness and linearity
analysis in the presence of rational trees. (Previous work
for rational tree unification has either focused
on pair-sharing \cite{K00} or set-sharing without freeness and linearity
\cite{HBZ02}).

In summary, this paper
provides the implementor with a
number of low-cost techniques for improving the precision and
efficiency of sharing analyses.

\section{Preliminaries}

\subsection{Trees and terms}

Let $\varepsilon$ denote the empty sequence, $.$ denote sequence
concatenation, and $\|\alpha\|$ denote the length of a sequence
$\alpha \in \nat^{*}$. A tree (or term) over an alphabet of
symbols $F$ is a partial map $t : \nat^{*} \to F$ such that
$t(\alpha) = t$ if $\alpha = \varepsilon$, otherwise $t(\alpha) =
{t_i}(\beta)$ where $\alpha = i . \beta$ and $t = f(t_1, \ldots,
t_n)$. Let $T(F)$ and $T^{\infty}(F)$ denote the set of finite and
possibly infinite trees over $F$. Let $U$ denote a (denumerable)
universe of variables such that $F \cap U = \emptyset$, and let
$var(t)$ = $\{ u \in U \mid \exists \alpha \in \nat^{*} .
t(\alpha) = u \}$ where $t \in T^{\infty}(F \cup U)$. Finally,
$|S|$ denotes the cardinality of the set $S$.

\subsection{Substitutions and equations}

A substitution is a (total) map $\theta : U \to T^{\infty}(F \cup
U)$ such that $dom(\theta)$ = \linebreak \mbox{$\{ u \in U \mid
\theta(u) \neq u \}$} is finite. A substitution $\theta$ can be
represented as a finite set $\{ x \mapsto \theta(x) \mid
x \in dom(\theta) \}$.  Let $rng(\theta)$ = $\cup \{
var(\theta(u)) \mid u \in dom(\theta) \}$ and let $Sub$ denote the
set of substitutions. If $\theta = \{ x_i \mapsto t_i
\}_{i=1}^{n}$ then $\theta(t)$ denotes the tree obtained by
simultaneously replacing each occurrence of $x_i$ in $t$ with
$t_i$. For brevity, let $\theta(x, \alpha)$ = $t(\alpha)$ where
$\theta(x)$ = $t$. An equation $e$ is a pair $(s = t)$ where $s, t
\in T^{\infty}(F \cup U)$. A finite set of equations is denoted $E$ and
$Eqn$ denotes the set of finite sets of equations. Also define
$\theta(E)$ = $\{ \theta(s) = \theta(t) \mid (s = t) \in E \}$.
The map $eqn : Sub \to Eqn$ is defined $eqn(\theta)$ = $\{ x = t
\mid (x \mapsto t) \in \theta \}$.  Where $Y \subseteq U$,
projection out and projection onto are respectively defined
$\exists Y . \theta$ = $\{ x \mapsto t \in \theta \mid x \not\in Y
\}$ and $\overline{\exists} Y . \theta$ = $\exists (U \setminus Y)
. \theta$. Composition $\theta \circ \psi$ of two substitutions is
defined so that $(\theta \circ \psi)(u) = \theta(\psi(u))$ for
all $u \in U$. Composition induces the (more general than)
relation $\leq$ defined by $\theta \leq \psi$ iff there exists
$\delta \in Sub$ such that $\psi = \delta \circ \theta$. A
renaming is a substitution $\rho \in Sub$ that has an inverse,
that is, there exists $\rho^{-1} \in Sub$ such that $\rho^{-1}
\circ \rho = id$. The set of renamings is denoted $Rename$. A
substitution $\theta$ is idempotent iff $\theta \circ \theta =
\theta$, or equivalently, iff $dom(\theta) \cap rng(\theta) =
\emptyset$.

\subsection{Solved forms and most general unifiers}

A substitution is in rational solved form iff it has no subset $\{
x_1 \mapsto x_2$, $\ldots$, $x_n \mapsto x_1 \}$ where $n \geq 2$.
The subset of $Sub$ in rational solved form is denoted $RSub$. The
set of unifiers of $E$ is defined by: $unify(E)$ = $\{ \theta \in
Sub \mid \forall (s = t) \in E . \theta(s) = \theta(t) \}$. The
set of most general unifiers (mgus) and the set of idempotent mgus (imgus)
are defined: $mgu(E) = \{ \theta
\in unify(E) \mid \forall \psi \in unify(E) . \theta \leq \psi \}$
and $imgu(E) = \{ \theta \in mgu(E) \mid dom(\theta) \cap
rng(\theta) = \emptyset \}$. Note that $imgu(E) \neq \emptyset$ iff
$mgu(E) \neq \emptyset$ \cite{LMM88}. An mgu can
be renamed to obtain any other (as can an imgu).

\begin{lemma}[\rm Proposition~11 from \cite{LMM88}]\label{lemma-lassez}
Let $\theta \in imgu(E)$. Then $\phi \in imgu(E)$ iff there exists
$\{ x_i \mapsto y_i \}_{i = 1}^{n} \subseteq \theta$ such that
$\phi = \{  x_i \mapsto y_i, y_i \mapsto x_i \}_{i = 1}^{n} \circ
\theta$.
\end{lemma}

\noindent One way to obtain an imgu is by considering
limits of substitutions.

\begin{definition}\label{defn-limits} \rm
Let $\{ t_{n} \mid n \in \nat \} \subseteq T^{\infty}(F \cup U)$.
Then $t = \lim_{n \to \infty} t_{n}$ iff for all $k \in \nat$
there exists $l \in \nat$ such that
for all $m \geq l$ and $\|\alpha\| \leq k$,
$t(\alpha) = {t_m}(\alpha)$.  Furthermore, if $\{
\theta_{n} \mid n \in \nat \} \subseteq Sub$ then $\lim_{n \to
\infty} \theta_{n} = \lambda x . \lim_{n \to \infty}
{\theta_{n}}(x)$.
\end{definition}

\noindent Note that $\lim_{n \to \infty} \theta^{n}$ exists iff
$\theta \in RSub$ \cite{K00}. Henceforth $\theta^\infty$
abbreviates $\lim_{n \to \infty} \theta^{n}$. If $\theta \in RSub$
then $\theta^\infty$ is idempotent whereas if $\theta$ is
idempotent then $\theta^\infty = \theta$. The following lemmas
detail how limits of substitutions and composition of
substitutions relate to an mgu.

\begin{lemma}[\rm Lemmas~2.2, 4.3 and 4.4 from \cite{K00}]\label{lemma-subs}
\rm
\begin{enumerate}

\item\label{case-one}
$\theta^{\infty} \in mgu(eqn(\theta))$ if $\theta \in RSub$.

\item\label{case-two}
$\delta \circ \theta^{\infty} \in mgu(E \cup eqn(\theta))$
if $\delta \in mgu({\theta^{\infty}}(E))$.

\item\label{case-three}
$\exists (dom(\theta) \setminus rng(\theta)) . \delta \in
mgu({\theta}(E))$
if $\delta \circ \theta \in mgu(E)$.

\end{enumerate}
\end{lemma}

\subsection{Linearity}

Variable multiplicity is defined in order to formalise
linearity. The significance of linearity is that
unification of linear terms enables sharing to
be described by more precise sharing
abstractions (even in the presence of rational trees).

\begin{definition} \rm
The variable multiplicity map $\chi : T^{\infty}(F \cup U) \to
\{0, 1, 2\}$ is defined: $\chi(t) = \max(\{ \chi(x, t) \mid x \in
U \})$ where $\chi(x, t) = \min(2, |\{ \alpha \mid t(\alpha) =
x\}|)$.
\end{definition}

\noindent If $\chi(t) = 0$, $t$ is ground; if $\chi(t) = 1$, $t$
is linear; and if $\chi(t) = 2$, $t$ is non-linear. The next
lemma details the forms of sharing barred by the unification of
linear terms.

\begin{lemma}[Proposition 3.1 from \cite{K00}]\label{lemma-original-linearity} \rm
If $\theta \in mgu(\{s = t\})$, $x \neq y$ and $var(\theta(x))
\cap var(\theta(y)) \neq \emptyset$ then either: $x \in var(s)$
and $y \in var(t)$; or $x, y \in var(t)$ and $\chi(s) = 2$; or $x
\in var(t)$ and $y \in var(s)$; or $x, y \in var(s)$ and $\chi(t)
= 2$.
\end{lemma}

\noindent The correctness arguments for abstract unification
require lemma~\ref{lemma-original-linearity} to be augmented with
a new result -- lemma~\ref{lemma-new-linearity}. The proof of this
lemma is analogous to that of lemma~\ref{lemma-original-linearity}
detailed in \cite{K00}.

\begin{lemma}\label{lemma-new-linearity} \rm
If $\theta \in mgu(\{s = t\})$ and $\chi(\theta(x)) = 2$ then
either: $x \in var(s) \cap var(t)$; or $x \in var(t)$ and $\chi(s)
= 2$; or $x \in var(s)$ and $\chi(t) = 2$.
\end{lemma}

\subsection{Groundness and sharing abstractions}

The abstract domains of interest in this paper
are represented either as Boolean functions, or
as sets or as sets of sets. Let $X$ denote a finite subset of $U$.
The set of
propositional formulae over $X$ is denoted by $\mybool_X$
and $Y$
abbreviates the formula $\wedge Y$.
The (bijective) map
${\mymodel_X} : \mybool_X \to \wp(\wp(X))$ is defined by
${\mymodel_X}(f) =
\{ M \! \subseteq \! X \! \mid \! {\psi_X}(M) \models f \}$
where
${\psi_X}(M) = M \land \land \{ \neg y \mid y \in X \! \setminus \! M \}$.
The groundness, sharing, freeness and linearity domains over $X$
are defined as follows:

\begin{definition}
$Pos_{X}$ = $\{ f \in Bool_X \mid X \models f \}$, $\Share_{X}$ =
$\{ S \subseteq \wp(X) \mid \emptyset \in S \}$, $\Free_{X}$ =
$\wp(X)$ and $\Lin_{X}$ = $\wp(X)$.
\end{definition}

\noindent If $S \in \Share_X$, then each $G \in S$ is referred
to as a sharing group.

These domains are connected to the concrete domain of
sets of equations by Galois connections induced
by the concretisation maps.
This approach leads to
succinct statements of correctness. To obtain well defined
concretisations,
maps abstracting substitutions are introduced. It is then
observed that the abstractions for
equivalent idempotent substitutions are the same.

\begin{definition} The abstraction maps $\alpha^{Pos} : Sub \to Pos_{U}$
and $\alpha^{\Share}_X : Sub \to \Share_{X}$ are defined:
$\alpha^{Pos}(\theta)$ = $\wedge \{ x \leftrightarrow var(t) \!
\mid \! x \mapsto t \in \theta \}$, $\alpha^{\Share}_X(\theta)$ =
$\{ occ(\theta, u) \cap X \mid u \in U \}$ and $occ(\theta, y) =
\{ u \in U \mid y \in var(\theta(u)) \}$.
\end{definition}

\begin{lemma} Let $\theta, \phi \in imgu(E)$. Then
$\alpha^{Pos}(\theta) = \alpha^{Pos}(\phi)$,
$\alpha^{\Share}_X(\theta) = \alpha^{\Share}_X(\phi)$, $\theta(x)
\in U$ iff $\phi(x) \in U$ and $\chi(\theta(x)) \leq 1$ iff
$\chi(\phi(x)) \leq 1$.
\end{lemma}

\begin{proof} By lemma~\ref{lemma-lassez} there exists
$\{ x_i \mapsto y_i \}_{i = 1}^{n} \subseteq \theta$ such that
$\phi = \rho \circ \theta$ where $\rho = \{  x_i \mapsto y_i, y_i
\mapsto x_i \}_{i = 1}^{n}$.
\begin{enumerate}

\item Let $x \mapsto t \in \theta$.
Observe that
$\{ x \mapsto \rho(t), y_1 \mapsto x_1, \ldots, y_n \mapsto x_n \} \subseteq \rho \circ \theta$ and $y_i \in var(t)$ iff $x_i \in var(\rho(t))$, thus
$\alpha^{Pos}(\phi) \models x \leftrightarrow var(\rho(t))
\wedge (\wedge_{i=1}^{n} y_i \leftrightarrow x_i) \models x
\leftrightarrow var(t)$. Hence $\alpha^{Pos}(\phi) \models
\alpha^{Pos}(\theta)$. The other direction is similar.

\item Observe that $occ(\rho \circ \theta, y_i) = occ(\theta, x_i)$,
$occ(\rho \circ \theta, x_i) = occ(\theta, y_i)$ and $occ(\rho
\circ \theta, u)$ = $occ(\theta, u)$ for all $u \not\in dom(\rho)
\cup rng(\rho)$. Hence $\alpha^{\Share}_X(\theta) =
\alpha^{\Share}_X(\phi)$.

\item and 4. Immediate.

\end{enumerate}
\end{proof}


\noindent Instead of defining concretisation in terms
of a particular imgu (the limit of
a rational solved form \cite{K00}), an
arbitrary imgu is used. This new approach simplifies
correctness proofs.

\begin{definition}\label{defn-concrete}
The concretisation maps \mbox{${\gamma^{Pos}_X} : Pos_X \to
\wp(Eqn)$}, \mbox{${\gamma^{\Share}_X} : \Share_X \to \wp(Eqn)$},
\linebreak \mbox{${\gamma^{\Free}_X} : \Free_X \to \wp(Eqn)$} and
\mbox{${\gamma^{\Lin}_X} : \Lin_X \to \wp(Eqn)$} are respectively
defined by:
\[ \begin{array}{r@{\; = \;}l}
{\gamma^{Pos}_X}(f) & \{ E \in Eqn \mid \exists \theta \in imgu(E)
. \alpha^{Pos}(\theta) \models f \}
\\
\gamma^{\Share}_{X}(S) & \{ E \in Eqn \mid \exists \theta \in
imgu(E) . \alpha^{\Share}_X(\theta) \subseteq S \}
\\
{\gamma^{\Free}_X}(F) & \{ E \in Eqn \mid \exists \theta \in
imgu(E) . \forall x \in F . \, \theta(x) \in U \}
\\
{\gamma^{\Lin}_X}(L) & \{ E \in Eqn \mid \exists \theta \in
imgu(E) . \forall x \in L . \chi(\theta(x)) \leq 1 \}
\end{array} \]
\end{definition}

\noindent Each free variable is linear so that
$\gamma^{\Free}_X(F) \cap \gamma^{\Lin}_X(L)$ =
$\gamma^{\Free}_X(F) \cap \gamma^{\Lin}_X(L \cup F)$. 
This paper
is concerned with combined domains and the following combined
concretisation maps will be useful: $\gamma_X^{SF}(\langle S, F
\rangle) = \gamma^{\Share}_X(S) \cap \gamma^{\Free}_X(F)$ and
$\gamma^{SFL}_X(\langle S, F, L \rangle)$ = $\gamma^{SF}_X(\langle
S, F \rangle) \cap \gamma^{\Lin}_X(L)$.

A connection is established in \cite{CSS99} which sheds light on
the relationship between sharing and Boolean functions.  The
corollary (also observed in the long version of \cite{BZH00})
explains how this can be used to improve precision of combined
domains.

\begin{lemma}[\rm Observation~4.1 and lemma 5.1 from
\cite{CSS99}]\label{lemma-codish} $\{X \setminus G \mid G\in
\alpha^{\Share}_X(\theta)\} \subseteq
model_X(\alpha^{Pos}(\theta))$ where $\theta$ is idempotent.
\end{lemma}

\begin{corollary}\label{cor-trim}
$\gamma^{Pos}_{X}(f) \cap \gamma^{\Share}_{X}(S) =
\gamma^{Pos}_{X}(f) \cap \gamma^{\Share}_{X}(trim_{X}(f, S))$
where $trim_X(f, S) = \{ G \in S \! \mid \! X \setminus G \in
model_{X}(f) \}$.
\end{corollary}

Finally, the following auxiliary operations will be
used throughout the paper. Let $S, S_i \in \Share_X$. The relevance map is defined
$rel(t, S) = \{ G \in S | var(t) \cap G \neq \emptyset \}$;
closure is defined $S^{*} = \cap \{ S' \mid S \subseteq S' \wedge
\forall G_1, G_2 \in S' . G_1 \cup G_2 \in S' \}$; and pair-wise
union is defined $S_1 \uplus S_2 = \{ G_1 \cup G_2 \mid G_1 \in
S_1 \wedge G_2 \in S_2 \}$.
Observe that if 
$var(rel(s, S)) \cap var(rel(t, S)) = \emptyset$
then $var(\theta(s)) \cap var(\theta(t)) = \emptyset$
for all $\theta \in imgu(E)$ and $E \in
\gamma^{\Share}_{X}(S)$. Thus the
independence check $var(rel(s, S)) \cap var(rel(t, S)) = \emptyset$
can verify that two
terms $s$ and $t$ do not
share under $\theta$ (or equivalently $E$).

\section{Independence check in set-sharing}

The following example demonstrates that pair-sharing can
sometimes detect independence when standard set-sharing
unification algorithms cannot.

\begin{example}\label{exam-independence} Let
$X = \{ u, v, w, x, y, z \}$ and consider $E \in
\gamma_X^{SFL}(\langle S, F, L \rangle)$ where $S = \{ \emptyset,
\{ u, w \}$, $\{ v, w \}$, $\{ x, y \}$, $\{ x, z \}$, $\{ w, x
\}\}$, $F = \emptyset$ and $L = X$. Let $\theta' \in imgu(E \cup
\{ w = x \})$. The set-sharing unification algorithms of
\cite{L91,BZH00} give the following abstraction $S' = \{ \emptyset
\} \cup (S_w^{*} \uplus S_x^{*})$ for $\theta'$ where $S_w = \{ \{
u, w \}, \{ v, w \}, \{ w, x \}\}$ and $S_x = \{ \{ x, y \}, \{ x,
z \}, \{ w, x \}\}$. Observe that $\{ u, v, w \} \in S_w^{*}$ and
$\{ x,y,z \} \in S_x^{*}$ and therefore $S'$ does not assert the
independence of $u$ and $v$ (similarly $y$ and $z$). However, if
$S$ is interpreted as a set of pairs, then the pair-sharing
abstract unification algorithms of \cite{CDY91,K00} both give the
abstraction $S \cup \{ \{ w \}, \{ x \}, \{ u, x \}, \{ u, y \},
\{ u, z \}, \{ v, x \}, \{ v, y \}, \{ v, z \}, \{ w, y \}, \{ w,
z \} \}$ which states the independence of $u$ and $v$ (and
similarly $y$ and $z$). Note that this different does not stem
from a difference in the set-sharing and pair-sharing
\textit{domains}, but derives from the way in which linearity is
exploited in the abstract unification \textit{algorithms}.
\end{example}

\noindent The crucial difference between pair-sharing and
set-sharing algorithms is that the former does not require the
terms in the equation to be independent to exploit linearity. Put
another way, to apply linearity the latter requires that
$var(rel(s, S)) \cap var(rel(t, S)) = \emptyset$ when solving
the equation $s = t$ in the context of the sharing abstraction $S$.
Lemmas~\ref{lemma-original-linearity}
and \ref{lemma-new-linearity} detail the forms of sharing that can
arise in $mgu(\{ s' = t' \})$ rational (and finite) tree
unification where $s'$ and $t'$ are arbitrary terms.
Observe that $s'$ and $t'$ are not required to be
independent. Abstract unification algorithms with the independence
check are safe. However, this check is not fundamental to
combining sharing with linearity. By observing how to exploit
linearity more fully, a more precise abstract unification
algorithm can be obtained. This algorithm also explains why
algorithms with the independence check are safe. The following
abstract operator is used to approximate the multiplicity map in
abstract unification. Lemma~\ref{lemma-chi} asserts its
correctness.

\begin{definition}
\[
\chi(t, S, L) = \left\{ \begin{array}{ll@{\,}c@{\,}r@{}l}
2 & \mathrm{if} \; \exists x & \in & var(S) & . \chi(x, t) = 2 \\
2 & \mathrm{if} \; \exists x & \in & var(S) & . x \in var(t) \setminus L \\
2 & \mathrm{if} \; \exists x, y & \in & var(t) & . \exists G \in S
. x \neq y \wedge x, y \in G \\
1 & \multicolumn{4}{l}{\mathrm{otherwise}}
\end{array} \right.
\]
\end{definition}

\begin{lemma}\label{lemma-chi} \rm If $E \in \gamma^{\Share}_X(S) \cap
\gamma^{\Lin}_X(L)$ and $\theta \in imgu(E)$ then $\chi(\theta(t))
\leq \chi(t, S, L)$.
\end{lemma}

\newpage

\begin{proof} Suppose $\chi(\theta(t)) = 2$. One of the
following holds:
\begin{itemize}

\item There exists $x \in var(t)$ such that $\chi(x, t) = 2$ and
$var(\theta(x)) \neq \emptyset$. Then \mbox{$x \in var(S)$} so
that $\chi(t, S, L) = 2$.

\item There exists $x \in var(t)$ such that $\chi(\theta(x)) = 2$.
Then $x \in var(S)$ and \linebreak $x \in var(t) \setminus L$ so
that $\chi(t, S, L) = 2$.

\item There exist $x, y \in var(t)$ such that $x \neq y$
and $var(\theta(x)) \cap var(\theta(y)) \neq \emptyset$. Then
there exists $G \in S$ such that $x, y \in G$ so that $\chi(t, S,
L) = 2$.

\end{itemize}
\end{proof}

\noindent The revised abstract unification algorithm (with the
independence check removed) is detailed in
definition~\ref{defn-amgu}, and
theorem~\ref{theorem-amgu-correct}
establishes its correctness.

\begin{definition}[Abstract unification 1]\label{defn-amgu}
Abstract unification
$amgu_1(\langle S, F, L \rangle, s, t) = \langle S', F', L' \rangle$
is defined:
\[
S_s = rel(s, S) \quad S_t = rel(t, S) \quad S' = (S \setminus (S_s
\cup S_t)) \cup S'' \quad G' = X \setminus var(S')
\]
\[
S'' = \left\{ \begin{array}{r@{\,}c@{\,}ll}
S_s & \uplus & S_t & \mathrm{if} \; s \in F \vee t \in F \\
(S_s^{*} \uplus S_t) & \cap & (S_s \uplus S_t^{*}) & \mathrm{if} \; \chi(s, S, L) = \chi(t, S, L) = 1 \\
S_s^{*} & \uplus & S_t & \mathrm{if} \; \chi(s, S, L) = 1 \\
S_s & \uplus & S_t^{*} & \mathrm{if} \; \chi(t, S, L) = 1 \\
S_s^{*} & \uplus & S_t^{*} & \mathrm{otherwise}
\end{array} \right.
\] \[
F' = \left\{ \begin{array}{ll}
F & \mathrm{if} \; s \in F \wedge t \in F \\
F \setminus var(S_s) & \mathrm{if} \; s \in F \\
F \setminus var(S_t) & \mathrm{if} \; t \in F \\
F \setminus var(S_s \cup S_t) & \mathrm{otherwise}
\end{array} \right.
\] \[
L' = F' \cup G' \cup \left\{ \begin{array}{ll} L \setminus
(var(S_s) \cap var(S_t)) & \mathrm{if} \; \chi(s, S, L) = 1
\wedge \chi(t, S, L) = 1 \\
L \setminus var(S_s) & \mathrm{if} \; \chi(s, S, L) = 1 \\
L \setminus var(S_t) & \mathrm{if} \; \chi(t, S, L) = 1 \\
L \setminus var(S_s \cup S_t) & \mathrm{otherwise}
\end{array} \right.
\]
\end{definition}

\noindent
A precision gain over
previous algorithms follows
since a closure is avoided if $s$ is linear but not $t$ (or vice versa) and
$s$ and $t$ are not independent.
When both $s$ and $t$ are linear, but
not independent, two closures are required (as
previously), but the resulting sharing
abstraction may contain fewer elements owing to the pruning
effect of intersection. When the independence check
is satisfied, that is
$S_s \cap S_t =
\emptyset$, it follows that
$(S_s^{*} \uplus S_t) \cap (S_s \uplus S_t^{*})$ = $S_s \uplus S_t$.
This explains why algorithms with the independence check are safe.
Note that if
$s$ and $t$ are both linear, but not independent, an
implementor might trade precision for efficiency
by computing
$S_s^{*} \uplus S_t$ if $|S_s| \leq |S_t|$ and
$S_s \uplus S_t^{*}$ otherwise.

\begin{theorem}[Correctness of abstract unification 1]\label{theorem-amgu-correct} \rm
Let $E \in \gamma_X^{SFL}(\langle S, F, L \rangle)$, $var(s) \cup
var(t) \subseteq X$ and $amgu_1(\langle S, F, L \rangle, s, t) =
\langle S', F', L' \rangle$. Then $E \cup \{ s = t \} \in
\gamma_X^{SFL}(\langle S', F', L' \rangle)$.
\end{theorem}

\begin{proof}
Put $E' = \{ s = t \}$.
Let $\theta \in imgu(E)$ and $\theta' \in imgu(E \cup E')$.
Observe that $unify(\theta(E')) \supseteq
unify(\theta(E') \cup eqn(\theta)) =
unify(E' \cup eqn(\theta)) =
unify(E \cup E') \neq \emptyset$. Thus let
$\delta \in imgu(\theta(E')) = imgu({\theta^{\infty}}(E'))$.
By part~\ref{case-two} of lemma~\ref{lemma-subs},
$\delta \circ \theta^{\infty} \in mgu(eqn(\theta) \cup E') = mgu(E \cup E')$.
Since $dom(\theta) \cap rng(\delta) = \emptyset$,
$\delta \circ \theta^{\infty} = \delta \circ \theta \in imgu(E \cup E')$.
\begin{enumerate}

\item To show $\alpha_X^{\Share}(\delta \circ \theta) \subseteq S'$,
let $y \in U$ and consider $occ(\delta \circ \theta, y)$.

\begin{enumerate}
\item Suppose $y \not\in rng(\delta \circ \theta)$.

\begin{enumerate}

\item Suppose $y \not\in dom(\delta \circ \theta)$, that is,
$\delta \circ \theta(y) = y$.  Thus
$\theta(y) = y'$ and $\delta(y')
= y$. Suppose $y \neq y'$.
Then $y \in dom(\theta)$, thus $y \not\in rng(\delta)$
which is a contradiction.
Therefore $y=y'$, giving $\theta(y) = y$ and
$\delta(y) = y$.

\begin{enumerate}

\item Suppose $y \not\in var(\theta(s))$ and $y \not\in var(\theta(t))$.
Hence $y \not\in dom(\delta)$ and \mbox{$y \not\in rng(\delta)$},
so that $occ(\delta \circ \theta, y) \cap X$ = $occ(\theta, y)
\cap X \in S$. But $var(s) \cap occ(\theta, y) = \emptyset$ and
similarly $var(t) \cap occ(\theta, y) = \emptyset$, so that
$occ(\delta \circ \theta, y) \cap X \in S'$.

\item
Suppose $y \in var(\theta(s))$ and $y \not\in var(\theta(t))$.
Since $\delta(y) = y$, it follows that $y \in var(\delta \circ
\theta(s)) = var(\delta \circ \theta(t))$. Suppose $y \in rng(\delta)$,
then $y \not\in dom(\theta)$, hence $y \in rng(\delta \circ \theta)$
which is a contradiction. Therefore $y \not\in rng(\delta)$, thus
$y \in var(\theta(t))$ which is a contradiction.

\item Suppose $y \not\in var(\theta(s))$ and $y \in var(\theta(t))$.
Analogous to the previous case.

\item Suppose $y \in var(\theta(s))$ and $y \in var(\theta(t))$.
Since $\delta(y)$ = $y$ and $y \not\in rng(\delta \circ \theta)$, $y
\not\in rng(\theta)$. Thus $y\in var(s)$ and $y\in var(t)$.
Since $y\not\in rng(\theta)$, it follows that $y \not\in
dom(\theta)$, therefore $y\not\in rng(\delta)$.
Thus, $occ(\delta\circ \theta, y) = occ(\theta, y)$.
Therefore $occ(\delta\circ \theta, y) \cap X \in S_s$ since $var(s)
\subseteq X$ and
$occ(\delta \circ \theta, y) \cap X \in S_t$ since $var(t) \subseteq X$.
Thus $occ(\delta \circ \theta, y) \cap X \in S'$.
\end{enumerate}


\item Suppose $y \in dom(\delta \circ \theta)$.
Since $y \not\in rng(\delta \circ \theta)$, $occ(\delta \circ
\theta, y) \cap X = \emptyset \in S'$.
\end{enumerate}

\item Suppose $y \in rng(\delta \circ \theta) \setminus var(\theta(E'))$.
Then $y \not\in dom(\delta)$ and
$y \not\in rng(\delta)$ so that $occ(\delta\circ \theta, y)
= occ(\theta, y)$. Moreover, since $y \not\in var(\theta(E'))$
it follows that $occ(\delta\circ \theta, y) \cap X \in S \setminus (S_s
\cup S_t)
\subseteq S'$.

\item Suppose $y \in rng(\delta \circ \theta) \cap var(\theta(E'))$.
Since $occ(\delta, y) \subseteq var(\theta(s)) \cup
var(\theta(t))$, $occ(\delta \circ \theta, y) \cap X = \cup \{
occ(\theta, u) \cap X \mid u \in occ(\delta, y) \} = (\cup R_s)
\cup (\cup R_t)$, where $R_s = \{ occ(\theta, v) \cap X \mid v \in
var(\theta(s)) \cap occ(\delta, y) \}$ and $R_t = \{ occ(\theta,
w) \cap X \mid w \in var(\theta(t)) \cap occ(\delta, y) \}$. 
If
$R_s = \emptyset$, then $y \not\in var(\delta \circ \theta(s)) =
var(\delta \circ \theta(t))$, hence $R_t = \emptyset$ and
$occ(\delta \circ \theta, y) \cap X = \emptyset \in S'$. Likewise
$occ(\delta \circ \theta, y) \cap X = \emptyset \in S'$ if $R_t =
\emptyset$. Thus suppose $R_s \neq \emptyset$ and $R_t \neq
\emptyset$. 
Since
$var(s) \subseteq X$, $R_s \subseteq S_s$ and since $var(t)
\subseteq X$, $R_t \subseteq S_t$.

\begin{enumerate}

\item\label{case-freeness}
Suppose $s \in F$. Thus $\theta(s) \in U$, hence $|R_s| =
|var(\theta(s))| = 1$. Moreover $\chi(\theta(s)) \leq 1$. 
Suppose $|R_t \setminus R_s| > 1$.
Thus there exists $u \neq v$ such that $u, v \in
var(\theta(t)) \setminus var(\theta(s))$ and
$var(\delta(u)) \cap var(\delta(v)) \neq
\emptyset$. This contradicts lemma~\ref{lemma-original-linearity},
hence $|R_t \setminus R_s| \leq 1$. Thus $occ(\delta \circ \theta, y)
\cap X \in S_s \uplus S_t$.

\item
Suppose $t \in F$. Analogous to the previous case.

\item
Suppose $\chi(s, S, L) = 1$. Thus $\chi(\theta(s)) \leq 1$. As
with case~\ref{case-freeness}, it follows that $|R_t \setminus R_s| \leq 1$.
Thus $occ(\delta \circ \theta, y) \cap X \in S_s^{*} \uplus S_t$.

\item
Suppose $\chi(t, S, L) = 1$. Analogous to the previous case.

\item
Otherwise $occ(\delta \circ \theta, y) \cap X \in S_s^{*} \uplus
S_t^{*}$.

\end{enumerate}

\end{enumerate}

\item It is straightforward to show
$\delta \circ \theta(x) \in U$ for all $x \in F'$.

%
%
%
%

\item To show $\chi(\delta \circ \theta(x)) \leq 1$ for all $x \in
L'$. Observe $\chi(\delta \circ \theta(x)) = 0$ if $x \in G'$ and
$\chi(\delta \circ \theta(x)) = 1$ if $x \in F'$. Hence, let $x
\in L \subseteq X$ and suppose $\chi(\delta \circ \theta(x)) = 2$.
\begin{enumerate}

\item Suppose $\chi(s, S, L) = 1$. By lemma~\ref{lemma-chi},
$\chi(\theta(s)) \leq 1$.
\begin{enumerate}

\item Suppose there exist $u, v \in var(\theta(x))$, $u \neq v$
such that $var(\delta(u)) \cap var(\delta(v)) \neq \emptyset$. By
lemma~\ref{lemma-original-linearity} either:
\begin{enumerate}

\item $u \in var(\theta(s))$ and $v \in var(\theta(t))$, hence $x
\in occ(\theta, u) \cap X \in S_s$, and therefore $x \not\in L'$.

\item $u \in var(\theta(t))$ and $v \in var(\theta(s))$, hence $x
\in occ(\theta, v) \cap X \in S_s$, and therefore $x \not\in L'$.

\item $u, v \in var(\theta(s))$. Hence $x
\in occ(\theta, v) \cap X \in S_s$, and thus $x \not\in L'$.

\end{enumerate}

\item Suppose there exists $u \in var(\theta(x))$ such that
$\chi(\delta(u)) = 2$. By lemma~\ref{lemma-new-linearity}, $u \in
var(\theta(s))$, thus $x \in occ(\theta, u) \cap X \in S_s$ and
therefore $x \not\in L'$.

\end{enumerate}

\item Suppose $\chi(t, S, L) = 1$. Analogous to the previous case.

\item Otherwise observe that either:
\begin{enumerate}

\item There exist $u, v \in var(\theta(x))$, $u \neq v$
such that $var(\delta(u)) \cap var(\delta(v)) \neq \emptyset$.
Thus \mbox{$u \in var(\theta(E'))$} and $x \in occ(\theta, u) \cap X \in
S_s \cup S_t$. Hence $x \not\in L'$.

\item There exists $u \in var(\theta(x))$ such that
$\chi(\delta(u)) = 2$. Thus \mbox{$u \in var(\theta(E'))$} and $x
\in occ(\theta, u) \cap X \in S_s \cup S_t$. Hence $x \not\in L'$.

\end{enumerate}
\end{enumerate}
\end{enumerate}
\end{proof}

\begin{example} Consider again example~\ref{exam-independence}.
Observe that $amgu_1(\langle S, F, L \rangle, w, x) = \langle S',
F', L' \rangle$ where $S' = \{ \emptyset \} \cup (S_w^{*} \uplus
S_x) \cap (S_w \uplus S_x^{*})$ = $\{ \emptyset, \{u,w,x\}$,
$\{u,w,x,y\}$, $\{u,w,x,z\}$, $\{v,w,x\}$, $\{v,w,x,y\}$,
$\{v,w,x,z\}$, $\{w,x\}$, $\{w,x,y\}$, $\{w,x,z\}\}$,
$F' = \emptyset$ and $L' = \emptyset$. This asserts the
independence of $u$ and $v$ (similarly $y$ and $z$), as required.
\end{example}

\noindent The following example, adapted from \cite{L91},
illustrates that closure can be required to abstract
the unification of linear terms.

\begin{example}\label{exam-check} Let
$X = \{w,x,y,z\}$ and observe $E \in \gamma_X^{SFL}(\langle S, F,
L \rangle)$ where $E = \{ w = f(x, y, z) \}$, $S = \{ \emptyset,
\{w,x\}, \{w,y\}, \{w,z\}\}$, $F = \emptyset$ and $L = \{w,x,y,z\}$.
Let $E' = \{ w = f(z, x, y) \}$ and note that $\theta' \in imgu(E
\cup E')$ where $\theta' = \{ w \mapsto f(z,z,z), x \mapsto z, y
\mapsto z \}$. Thus $E \cup E' \in \gamma_X^{SFL}(\langle S', F',
L' \rangle)$ where $S' = \{ \emptyset, \{w,x,y,z\}\}$, $F' =
\emptyset$ and $L' = \{x,y,z\}$. Indeed, if $S_s = rel(w, S) =
\{\{w,x\}, \{w,y\}, \{w,z\}\}$ and $S_t = rel(f(z, x, y), S) =
\{\{w,x\}, \{w,y\}, \{w,z\}\}$ then $(S_s^{*} \uplus S_t) \cap
(S_s \uplus S_t^{*})$ = $\{\{w,x\}$, $\{w,y\}$, $\{w,z\}$,
$\{w,x,y\}$, $\{w,x,z\}$, $\{w,y,z\}$, $\{w,x,y,z\}\}$, thus
$amgu_1(\langle S, F, L\rangle$, $w, f(z,x,y))$
yields a safe, though conservative, abstraction. Closure is
required to construct the $\{w,x,y,z\}$ sharing group.
\end{example}

\section{Decomposition of set-sharing}

Fil\'{e} \cite{F94} observes that different sharing and freeness
abstractions can represent the same equations, that is,
$\gamma^{SF}_X(\langle S_1, F \rangle) = \gamma^{SF}_X(\langle
S_2, F\rangle)$ does not imply that $S_1 = S_2$. Therefore the
relationship between $\Share \times \Free$ and the concrete domain is a
Galois connection rather than an insertion. An insertion is
constructed by using $F$ to decompose $S$ into a set of sharing
abstractions $K_F(S)$ such that each $B \in K_F(S)$ does not
include sharing groups that definitely arise from different
computational paths. The following definition and lemma from \cite{F94}
formalises this decomposition, henceforth referred to as the
Fil\'{e} decomposition.

\begin{definition}
The map $K_F(S) : \Share \to \wp(\Share)$ is defined by:
\[
K_F(S) = \left\{ B \left|
\begin{array}{c}
B \subseteq S \qquad \wedge \qquad F \subseteq var(B) \qquad \wedge \\
\forall G_1, G_2 \in B . (G_1 \neq G_2 \rightarrow G_1 \cap G_2
\cap F = \emptyset)
\end{array}
\right. \right\}
\]
\end{definition}

\begin{lemma}\label{lemma-k} $\gamma^{SF}_X(\langle S, F \rangle) =
\cup \{ \gamma^{SF}_X(\langle B, F \rangle) \mid B \in K_F(S) \}$.
\end{lemma}

\noindent Using the above, abstract unification can be refined to
$\cup \{ amgu(\langle B, F, L \rangle, s, t) | B \in K_F(S)\}$.
Abstract unification computed in this way does not merge sharing groups arising
from different computational paths, and thereby improves precision.
Calculating $K_F(S)$ is expensive and
the number of calls to $amgu$ is $|K_F(S)|$ (which is
potentially exponential in $|S|$).
However, this tactic suggests lightweight refinements to
closure ($*$) and pair-wise union ($\uplus$) that recover some precision
at little cost.
Since two distinct sharing groups which contain a common free variable
must arise from different computational paths, they cannot describe
the same equation and therefore
need not be combined.  Definition~\ref{defn-amgu2} details the
refined abstract unification algorithm and theorem~\ref{cor-amgu2}
builds on lemma~\ref{lemma-fiddly} to establish correctness.

\begin{definition}[Abstract unification 2]\label{defn-amgu2}
Abstract unification $amgu_{2}(\langle S, F, L \rangle, s, t) =
\langle S', F', L' \rangle$ is defined:
\[
S'' = \left\{ \begin{array}{r@{\,}l@{\,}ll}
(S_s^{*_{F}} \uplus_F S_t) & \cap & (S_s \uplus_F S_t^{*_{F}}) & \mathrm{if} \; \chi(s, S, L) = \chi(t, S, L) = 1 \\
S_s^{*_{F}} & \uplus_F & S_t & \mathrm{if} \; \chi(s, S, L) = 1 \\
S_s \,\,\,\, & \uplus_{F} & S_t^{*_{F}} & \mathrm{if} \; \chi(t, S, L) = 1 \\
S_s^{*_{F}} & \uplus_{F} & S_t^{*_{F}} & \mathrm{otherwise}
\end{array} \right.
\]
\[
S_1 \uplus_F S_2 = \bigcup \left\{ G_1 \cup G_2 \left|
\begin{array}{@{}c@{}}
G_1 \in S_1 \wedge G_2 \in S_2 \wedge 
G_1 \neq G_2 \rightarrow G_1 \cap G_2 \cap F = \emptyset
\end{array} \right. \right\}
\]
\[
S^{*_F} = \bigcap \left\{ S' \left|
\begin{array}{@{}c@{}}
S \subseteq S' \wedge \forall G_1, G_2 \in S' .
G_1 \cap G_2 \cap F = \emptyset \rightarrow G_1 \cup G_2 \in S'
\end{array} \right. \right\}
\]
where $S'$, $S_s$, $S_t$, $F'$ and $L'$ are defined as in
definition~\ref{defn-amgu}.
\end{definition}

\noindent Notice that the use of freeness is completely
absorbed into $*_{F}$ and $\uplus_{F}$. The following lemma demonstrates that $\uplus_F$
and $*_F$ coincide with $\uplus$
and $*$ for each element of the Fil\'{e} decomposition. The correctness
of abstract unification ($amgu_2$) follows from this result.

\begin{lemma}\label{lemma-fiddly}
\begin{enumerate}
\item If $B \in K_F(S)$ and $R \subseteq B$,
then $R^{*} = R^{*_F}$.

\item If $B \in K_F(S)$ and $R_1, R_2 \subseteq B$,
then $R_1 \uplus R_2 = R_1 \uplus_{F} R_2$, $R_1^{*} \uplus R_2 =
R_1^{*} \uplus_{F} R_2$, $R_1 \uplus R_2^{*} = R_1 \uplus_{F}
R_2^{*}$ and $R_1^{*} \uplus R_2^{*} = R_1^{*} \uplus_{F}
R_2^{*}$.
\end{enumerate}
\end{lemma}

\begin{proof}
\begin{enumerate}

\item Proof by induction.
\begin{enumerate}

\item Suppose $R = \emptyset$. Then $R^{*} = \emptyset = R^{*_F}$.

\item Suppose $R = \{ G \} \cup R'$. By the hypothesis,
${R'}^{*} = {R'}^{*_F}$. Since $R \subseteq B$,
then for all $G' \in R'$, $G' \cap G \cap F = \emptyset$.
Hence $R^{*} = R^{*_F}$.

\end{enumerate}

\item \begin{enumerate}

\item To show $R_1 \uplus R_2 = R_1 \uplus_{F} R_2$. Let $G_i \in
R_i$. If $G_1 \cap G_2 \cap F \neq \emptyset$ then $G_1 = G_2$.
Hence $G_1 \cup G_2 \in R_1 \uplus_{F} R_2$.

\item To show $R_1^{*} \uplus R_2 = R_1^{*} \uplus_{F} R_2$.
Let $G_1 \in R_1^{*}$ and $G_2 \in R_2$. Then $G_1 = \cup Q_1$ for
some $Q_1 \subseteq R_1$. Put $Y = G_1 \cap G_2 \cap F$, $Q_1' =
\{ G \in Q_1 \mid G \cap Y = \emptyset \}$ and $Q_1'' = Q_1
\setminus Q_1'$. Observe that $|Q_1''| \leq 1$ and $Q_1''
\subseteq \{ G_2 \}$. Thus $G_1 \cup G_2 = (\cup Q_1') \cup G_2$.
Since $(\cup Q_1') \cap G_2 \cap F = \emptyset$ it follows that
$G_1 \cup G_2 \in R_1^{*} \uplus_{F} R_2$.

\item To show $R_1 \uplus R_2^{*} = R_1 \uplus_{F} R_2^{*}$. Analogous
to the previous case.

\item To show $R_1^{*} \uplus R_2^{*} = R_1^{*} \uplus_{F} R_2^{*}$.
Let $G_1 \in R_1^{*}$ and $G_2 \in R_2^{*}$. Then $G_i = \cup Q_i$
for some $Q_i \subseteq R_i$. Put $Y = G_1 \cap G_2 \cap F$, $Q_i'
= \{ G \in Q_i \mid G \cap Y = \emptyset \}$ and $Q_i'' = Q_i
\setminus Q_i'$. Observe that $|Q_i''| \leq 1$.
\begin{enumerate}

\item Suppose $|Q_1''| = \emptyset$ or $|Q_2''| = \emptyset$. Then
$G_1 \cap G_2 \cap F =
\emptyset$, hence $G_1 \cup G_2 \in R_1^{*} \uplus_{F} R_2^{*}$.

\item Suppose $|Q_1''| = |Q_2''| = 1$. Hence $Q_1'' = Q_2''$,
thus $G_1 \cup G_2 = G_1 \cup (\cup Q_2')$. Since $G_1 \cap (\cup
Q_2') \cap F = \emptyset$ it follows that $G_1 \cup G_2 \in
R_1^{*} \uplus_{F} R_2^{*}$.

\end{enumerate}
\end{enumerate}
\end{enumerate}
\end{proof}

\begin{theorem}[Correctness of abstract unification 2]\label{cor-amgu2} \rm
Let $E \in \gamma_X^{SFL}(\langle S, F, L \rangle)$, $var(s) \cup
var(t) \subseteq X$ and $amgu_{2}(\langle S, F, L \rangle, s, t) =
\langle S', F', L' \rangle$. Then $E \cup \{ s = t \} \in
\gamma_X^{SFL}(\langle S', F', L' \rangle)$.
\end{theorem}

\begin{proof} Observe $E \in \gamma_X^{SF}(\langle S, F \rangle)$
and $E \in \gamma_X^{L}(L)$. By lemma~\ref{lemma-k}, there exists
$B \in K_F(S)$ such that $E \in \gamma_X^{SF}(\langle B, F
\rangle)$, hence $E \in \gamma_X^{SFL}(\langle B, F, L \rangle)$.
Observe that if $s \in F$ then $S_s^{*_F} = S_s$
(and likewise for $t \in F$)
and hence by lemma~\ref{lemma-fiddly}, $amgu_1(\langle B, F, L \rangle, s, t)
= amgu_{2}(\langle B, F, L \rangle, s, t)$.
By theorem~\ref{theorem-amgu-correct}, $E \cup \{ s = t
\} \in \gamma_X^{SFL}(amgu_1(\langle B, F, L \rangle, s, t))$
= $\gamma_X^{SFL}(amgu_2(\langle B, F, L \rangle, s, t))$,
thus $E \cup \{ s = t
\} \in \gamma_X^{SFL}(amgu_2(\langle S, F, L \rangle, s, t))$.
\end{proof}

\noindent The proof explains why
the standard freeness tactic is a specialised
version of the Fil\'{e} decomposition.

This refinement is only
worthwhile if redundant sharing groups are introduced in analysis.
Although it can be shown that projection and join do not introduce
redundancy, the following example indicates that redundant sharing
groups can arise in abstract unification ($amgu_1$)
and that the refined abstract unification ($amgu_2$) can avoid
some of these redundant sharing groups.

\newpage

\begin{example}\label{exam-redundant} Let $X = \{ x, y, z \}$, $S = \{\emptyset, \{x, y\}, \{ y, z \}\}$,
$F = \{ y \}$ and $L = \{ y \}$. Suppose $s= x$ and $t=z$. Then
$S_s = \{ \{ x, y \} \}$ and $S_t = \{ \{ x, z \} \}$ so that
$amgu_1(\langle S, F, L \rangle, x, z) = \langle \{ \emptyset, \{
x, y, z \} \}, \emptyset, \emptyset \rangle$. However ${S_s}^{*_F}
= \{ \{ x, y \} \}$ and ${S_t}^{*_F} = \{ \{ x, z \} \}$ and in
particular ${S_s}^{*_F} \uplus_{F} {S_t}^{*_F} = \emptyset$ so
that $amgu_2(\langle S, F, L \rangle, x, z)$ = $\langle  \{
\emptyset \}, \emptyset, \{ x, y, z \} \rangle$.
\end{example}

\noindent The following example demonstrates that $amgu_2$ is not
as precise as the full Fil\'{e} decomposition.

\begin{example} Let $X = \{x,y,z\}$,
$S = \{ \emptyset, \{x\}, \{z\}, \{x,y\}, \{y,z\}\}$, $F = \{x, y,
z\}$ and $L = \{x, y, z\}$. Suppose $s= x$ and $t=z$. Then ${S_s}
= \{ \{ x \}, \{ x, y \} \}$ and ${S_t} = \{ \{ z \}, \{ y, z \}
\}$, hence ${S_s}^{*_F} = {S_s}$ and ${S_t}^{*_F} = {S_t}$. Thus
${S_s}^{*_F} \uplus_F {S_t}^{*_F}$ = $\{ \emptyset, \{x, z\},
\{x,y,z\}\}$.  It follows that $amgu_2(\langle S, F, L \rangle, x,
z)$ = $\langle \{ \emptyset, \{x, z\}, \{x,y,z\}\}, F, L \rangle$.
However, the Fil\'{e} decomposition gives $K_F(S) = \{ S_1, S_2,
S_3, S_4 \}$ where $S_1 = \{ \{x\}, \{y,z\} \}$, $S_2 = \{
\emptyset, \{x\}, \{y,z\} \}$, $S_3 = \{ \{x,y\}, \{z\} \}$ and
$S_4 = \{ \emptyset, \{x,y\}, \{z\} \}$. Moreover, $amgu_1(\langle
S_2, F, L \rangle, x, z)$ = $amgu_1(\langle S_4, F, L \rangle, x,
z)$ = $\langle \{ \emptyset, \{x,y,z\}\}, F, L \rangle$. Since
$S_1 \subseteq S_2$ and $S_3 \subseteq S_4$, the Fil\'{e} leads to
the sharing abstraction $\{ \emptyset, \{x,y,z\}\}$, which is more
precise.
\end{example}

\section{Pruning of set-sharing}

Pruning sharing groups is advantageous for efficiency and
precision. By reducing the size of an abstraction, abstract
unification works on smaller objects and is therefore faster,
even if no precision is gained. Of course, the benefit
of pruning for efficiency needs to outweigh its cost.

\subsection{Pruning with freeness via groundness}

Surprisingly, combined
sharing and freeness information can
improve groundness propagation and
sharing even for rational tree unification. For example, the equation \mbox{$x = f(y, z)$}
can be abstracted by $(x \leftrightarrow z) \wedge (x \leftrightarrow y)$
if $x$ and $y$ are free variables that share. This is because,
in this circumstance, finite tree
unification fails for $x = f(y, z)$
whereas rational tree
unification binds $x$ and $y$ to $f(f(\ldots, z), z)$.
Abstract unification can use the freeness of variables
in the equation to extract hidden groundness information (for
distinct computational paths)
and thereby prune sharing groups and improve precision. The proof of theorem~\ref{cor-amgu3}
again uses the Fil\'{e} decomposition.

\begin{definition}[Abstract unification 3]\label{defn-amgu3}
Abstract unification $amgu_{3}(\langle S, F, L \rangle, s, t) =
\langle S', F', L' \rangle$ is defined:
\[ S' = (S \setminus (S_s \cup S_t))
\cup \left\{ \begin{array}{@{}cl@{}} \bigcup_{G \in S_s}
trim_X(s \leftrightarrow Y, \{ G \} \uplus_{F} \!\! S_t)
& \mbox{if} \; s \in F \wedge t \not\in U \\
\bigcup_{G \in S_t} trim_X(Z \leftrightarrow t , S_s
\uplus_{F} \!\! \{ G \}) & \mbox{if} \; t \in F \wedge s \not\in U \\
S'' & \mbox{otherwise}
\end{array} \right.
\]
where $Y = var(t) \setminus (G \cap F)$, $Z = var(s) \setminus (G
\cap F)$, $S_s$, $S_t$, $S''$, $F'$ and $L'$ are defined as in
definition~\ref{defn-amgu2}.
\end{definition}

\begin{theorem}[Correctness of abstract unification 3]\label{cor-amgu3} \rm
Let $E \in \gamma_X^{SFL}(\langle S, F, L \rangle)$, $var(s) \cup
var(t) \subseteq X$ and $amgu_{3}(\langle S, F, L \rangle, s, t) =
\langle S', F', L' \rangle$. Then $E \cup \{ s = t \} \in
\gamma_X^{SFL}(\langle S', F', L' \rangle)$.
\end{theorem}

\begin{proof}
Suppose $s \in F$. By lemma~\ref{lemma-k}, there exists $B \in
K_F(S)$ such that \mbox{$E \in \gamma_X^{SFL}(\langle B, F
\rangle)$} and by theorem~\ref{cor-amgu2}, $E \cup \{ s = t \}
\in \gamma^{\Share}_X(B')$ where $B' = (B \setminus (B_s \cup
B_t)) \cup (B_s \uplus_{F} B_t)$, $B_s = rel(s, B)$ and $B_t =
rel(t, B)$. Let $\theta \in imgu(E)$. Since $s \in F$, $\theta(s)
= x$ for some $x \in U$. Furthermore, $s \in G$ for all $G \in
S_s$. Since $s \in F$, $B_s = \{ G \}$ where $G = occ(\theta, x)$.
Observe that $\theta(y) = x$ for all $y \in G \cap F$. Since $t
\not\in U$, $\theta(t) \not\in U$, hence $\alpha_X^{Pos}(\{
\theta(s) = \theta(t) \}) \models s \leftrightarrow Y$. Moreover,
$mgu(E \cup \{ s = t \}) = mgu(eqn(\theta) \cup \{ s = t \}) =
mgu(eqn(\theta) \cup \{ \theta(s) = \theta(t) \})$. Thus
$\alpha_X^{Pos}(E \cup \{ s = t \}) \models \alpha_X^{Pos}(\{
\theta(s) = \theta(t) \}) \models s \leftrightarrow Y$. The result
follows by corollary~\ref{cor-trim}. The $t \in F$ case is
analogous and the otherwise case follows immediately from
theorem~\ref{cor-amgu2}.
\end{proof}

\noindent The following example illustrates the gain of precision.
Note that even the Fil\'{e} decomposition cannot match this level
of precision.

\begin{example} \rm Let $X = \{ x, y, z \}$,
$S = \{ \emptyset, \{x,y\}, \{y\}, \{z\} \}$, $F = \{ x, y \}$ and
$L = \{ x, y \}$. \linebreak Suppose $s = x$ and $t = f(y, z)$.
Consider the Fil\'{e} decomposition, that is, $K_F(S) = \{ S_1,
S_2, S_3, S_4 \}$ where $S_1 = \{ \{ x, y \} \}$, $S_2 = \{
\emptyset, \{ x, y \} \}$, $S_3 = \{ \{ x, y \}, \{ z \} \}$, \linebreak $S_4
= \{ \emptyset, \{ x, y \}, \{ z \} \}$. Then $amgu_{1}(\langle
S_4, F, L \rangle, x, f(y, z))$ = $\langle S', \emptyset,
\emptyset \rangle$ where $S' = \{ \emptyset$, $\{x,y\}$,
$\{x,y,z\} \}$. Since 
$S_i \subseteq S_4$
for all $i \in \{ 1, 2, 3 \}$, the decomposition
results in the sharing abstraction $S'$. 
Moreover, $amgu_{2}(\langle S, F, L \rangle, x, f(y, z))$ = $\langle S',
\emptyset, \emptyset \rangle$. 
However,
$amgu_{3}(\langle S, F, L \rangle, x, f(y, z))$ = $\langle
trim_X(x \leftrightarrow z, S'), \emptyset, \emptyset \rangle$ =
$\langle \{ \emptyset, \{x,y,z\} \}, \emptyset,
\emptyset \rangle$ which is more precise.
\end{example}

\begin{example}\label{exam-optimal} Let $X = \{ x, y, z \}$,
$S = \{\emptyset, \{x, y\}, \{ y, z \}\}$, $F = \{ y \}$ and $L =
\{ y \}$. Suppose \linebreak $s= x$ and $t=z$. Since $x, z \in U$,
$amgu_3(\langle S, F, L \rangle, x, z)$ = $amgu_2(\langle S, F, L
\rangle, x, z)$ = $\langle \{ \emptyset, \{ x, y, z \} \},
\emptyset, \emptyset \rangle$ whereas the Fil\'{e} decomposition
produces $\langle \{ \emptyset \}, \emptyset, \{ x, y, z \}
\rangle$ \linebreak (see example~\ref{exam-redundant}).
\end{example}

\noindent Example~\ref{exam-optimal} shows that $amgu_3$ is not
uniformly more precise than the Fil\'{e} decomposition, hence is
sub-optimal. Nevertheless, this pruning tactic suggests that any
optimal abstract unification algorithm for sharing, freeness and
linearity, in the presence of groundness, will have to consider
subtle interactions between the components.

\subsection{Early pruning with groundness}

Sharing abstractions can always be pruned by removing sharing
groups which contain ground variables. Common practice is to
schedule the solving of equations so as to first apply abstract
unification to equations on ground terms \cite{L91}. Moreover,
\cite{MH92} details a
queueing/dequeueing mechanism for maximally
propagating groundness among systems of equations. This can
involve repeated searching. This section proposes a revision of
this tactic that applies groundness to the complete set of
equations (without repeated searching) and then uses the resulting
groundness information to prune sharing before abstract
unification is applied. The gain is that searching and scheduling
are no longer required (the mechanism is single pass) and that the
disjunctive groundness information captured by $Pos$ can be
exploited so that abstract unification can potentially operate on
smaller abstractions. Observe that groundness information will
normally be tracked by $Pos$ anyway, thus the computational
overhead is negligible. To formulate this strategy, abstract
unification is lifted to sets of equations as follows:

\begin{definition} The map
$amgu_{i}(T, E) = \{ T' \mid \langle T, E \rangle
\rightsquigarrow^{\star} \langle T', \emptyset \rangle \}$ is
defined by the least relation $\rightsquigarrow \, \subseteq
(Share_{X} \times \Free_X \times Lin_X)^{2}$ such that $\langle T,
\{ s = t \} \cup E \rangle \rightsquigarrow \langle amgu_{i}(T, s,
t), E \rangle$.
\end{definition}

\noindent The following theorem states correctness of the early
pruning using groundness for $amgu_1$, $amgu_2$ and $amgu_3$.

\begin{theorem}\label{theorem-ground-safety} \rm
Let $E \in \gamma^{Pos}_X(f) \cap \gamma^{SFL}_X(\langle S, F, L
\rangle)$, $E \cup E' \in \gamma^{Pos}_X(f')$, \mbox{$Y = \{ y \in X
\mid f' \models y \}$}, \mbox{$S' = trim_X(f \wedge Y, S)$},
\mbox{$F' = F \setminus var(rel(Y, S))$}, \mbox{$L' = L \cup Y$},
$var(E) \subseteq X$
and \linebreak \mbox{$T' \in amgu_{i}(\langle S', F', L' \rangle, E')$}.
Then $E \cup E' \in \gamma^{SFL}_X(T')$.
\end{theorem}

\begin{proof} \rm
Let $\theta \in imgu(E)$ and $\theta' \in imgu(E \cup E')$. Since
$\theta' \in unify(E)$, $\theta \leq \theta'$ and there exists
$\zeta \in Sub$ such that $\zeta \circ \theta = \theta'$. Since
$\theta' \in unify(E')$, $\zeta \in unify(\theta(E'))$ so that
$mgu(\theta(E')) \neq \emptyset$. Let $\delta \in
imgu(\theta(E')) = imgu(\theta^{\infty}(E'))$.
By part~\ref{case-two} of lemma~\ref{lemma-subs},
$\delta \circ \theta = \delta \circ \theta^{\infty} \in
mgu(eqn(\theta) \cup E') = mgu(E \cup E')$.
Thus there exists $\rho \in Rename$ such that
$\rho \circ \delta \circ \theta = \theta'$.
Now $var(\theta'(y)) = \emptyset$ for all $y \in Y$, hence
$var(\delta \circ \theta(y)) = \emptyset$ for all $y \in Y$.
Put $Z = \cup \{ var(\theta(y)) \mid
y \in Y \}$, $\phi = \overline{\exists}Z .
\delta$ and $\psi = \exists Z . \delta$.
Let $z \in Z$. Then there exists $y \in Y$ such that
$z \in var(\theta(y))$. But $var(\delta \circ \theta(y)) = \emptyset$,
hence $rgn(\phi) = \emptyset$ and $\delta =
\psi \circ \phi$. Thus $\psi \circ \phi \in mgu(\theta(E'))$ and
by lemma~\ref{lemma-subs} part~\ref{case-three},
${\exists} (dom(\phi) \setminus rng(\phi)) . \psi \in
mgu(\phi \circ \theta(E'))$. Furthermore, $\exists(dom(\phi)
\setminus rng(\phi)) . \psi = \psi$ hence $\psi \in mgu(\phi
\circ \theta(E'))$. Since $\phi \circ \theta$ is idempotent,
\mbox{$\psi \in mgu((\phi \circ \theta)^{\infty}(E'))$}. By
lemma~\ref{lemma-subs}, part~\ref{case-two}, $\psi \circ \phi
\circ \theta = \psi \circ (\phi \circ \theta)^{\infty} \in
mgu(eqn(\phi \circ \theta) \cup E')$.
Thus $\theta' \in imgu(eqn(\phi \circ \theta) \cup E')$.

To show
$eqn(\phi \circ \theta) \in \gamma_X^{\Share}(trim(f \wedge Y, S))$.
Let $u \in U$. If $occ(\phi \circ \theta, u) = \emptyset$ then $occ(\phi
\circ \theta, u) \cap X \in S$ trivially. If $occ(\phi \circ
\theta, u) \neq \emptyset$ then $occ(\phi \circ \theta, u) =
occ(\theta, u)$ since $rng(\phi) = \emptyset$.  Thus $occ(\phi
\circ \theta, u) \cap X \in S$. Therefore $eqn(\phi \circ \theta)
\in \gamma_X^{\Share}(S)$. By lemma~\ref{lemma-subs},
part~\ref{case-two}, $\delta \circ \theta \in mgu(E \cup
eqn(\theta))$. But $\theta' \in mgu(E \cup eqn(\theta))$ and
therefore there exists $\rho \in Rename$ such that $\rho \circ
\delta \circ \theta = \theta'$. Thus $\alpha^{Pos}(\delta \circ
\theta) \models \alpha^{Pos}(\rho \circ \delta \circ \theta) =
\alpha^{Pos}(\theta') \models Y$. Observe that if
$\alpha^{Pos}(\delta \circ \theta) \models u$ then
$\alpha^{Pos}(\phi \circ \theta) \models u$ hence
$\alpha^{Pos}(\phi \circ \theta) \models Y$. Since
$\alpha^{Pos}(\phi \circ \theta)  \models \alpha^{Pos}(\theta)
\models f$, it follows that $\alpha^{Pos}(\phi \circ \theta)
\models f\wedge Y$. Therefore $eqn(\phi \circ \theta) \in
\gamma_X^{Pos}(f \wedge Y)$. By corollary~\ref{cor-trim},
$eqn(\phi \circ \theta) \in \gamma_X^{\Share}(trim(f \wedge Y, S))$.

To show $\phi \circ \theta(x) \in U$ for all $x \in F'$.
Let $x \in F$ and $x \not\in var(rel(Y, S))$.
Since $x \not\in var(rel(Y, S))$, $x \not\in occ(\theta, u) \cap X$
or $y \not\in occ(\theta, u) \cap X$ for all $u \in U$ and $y \in Y$.
Since $x \in X$ and $Y \subseteq X$,
$var(\theta(x)) \cap var(\theta(y)) = \emptyset$ for all $y \in Y$.
Hence $\theta(x) \not\in Z$, thus $\theta(x) \not\in dom(\phi)$,
therefore $\phi \circ \theta(x) \in U$.
Thus $eqn(\phi \circ \theta) \in \gamma_X^{\Free}(F')$.

To show $\chi(\phi \circ \theta(x)) \leq 1$ for all $x \in L'$.
Since $rng(\phi) = \emptyset$, $\chi(\phi \circ \theta(x)) \leq 1$
for all $x \in L$. Moreover, $\alpha^{Pos}(\phi \circ \theta) \models Y$
and therefore $\chi(\phi \circ \theta(x)) \leq 1$
for all $x \in Y$.
Thus $eqn(\phi \circ \theta) \in \gamma_X^{\Lin}(L')$.
The result then follows by induction on $E$ and
theorems~\ref{theorem-amgu-correct}, \ref{cor-amgu2} and \ref{cor-amgu3}.
\end{proof}

\noindent The following example illustrates the
computational advantages of early pruning.

\begin{example} \rm
Let $X = \{u,v,x,y\}$,
$S = \{ \emptyset, \{x\}, \{y\}, \{u\}, \{v\} \}$,
$F = \emptyset$,
$L = \emptyset$ and
$f = x \vee y$. Let $E' = \{ x = f(u, v), x = y \}$ so
that $f' = (x \vee y) \wedge
(x \leftrightarrow (u \wedge v)) \wedge
(x \leftrightarrow y)$ = $x \wedge y \wedge u \wedge v$. Then
$Y = \{ x, y, u, v \}$ so that $f \wedge Y = x \wedge y \wedge u \wedge v$
and $S' = trim_X(f \wedge Y, S)$ = $\{ \emptyset \}$.
Hence $amgu_3(\langle S, F, L \rangle, E')$
reduces to $amgu_3(\langle S', F, L \rangle, E')$ =
$\langle \{ \emptyset \}, \emptyset, \emptyset \rangle$. Without
this tactic, no equation of $E'$ will possess a ground
argument and both calls to $amgu_3$ will involve non-trivial
sharing group manipulation.
\end{example}

\section{Conclusion}

This paper has given correctness proofs for sharing analysis with
freeness and linearity which hold in the presence of rational
trees. The abstract unification algorithms are themselves novel --
incorporating optimisations for both precision and efficiency.
Specifically, the independence check which can prevent linearity
from being exploited has been removed. In addition, refined
closure and pair-wise union operations have been derived from the
Fil{\'e} decomposition. 
A further precision optimisation has been presented which exploits
an interaction between sharing, freeness and groundness, which
shows the subtlety that an optimal algorithm will need to address.
These optimisations have been chosen to balance precision against
efficiency whilst not changing the underlying representation of the
abstract domains. They are
ordered according to 
their anticipated degree of usefulness.
This work provides the implementor with a suite of new
optimisations for abstract unification algorithms for sharing,
freeness and linearity.

\subsubsection*{Acknowledgements}
We thank Gilberto~Fil\'{e} for kindly sending us a copy
of his technical report.
This work was supported, in part, by EPSRC grant
GR/MO8769.


\begin{thebibliography}{}

\bibitem[\protect\citename{Bagnara {\em et~al.}\relax, }2000]{BZH00}
Bagnara, R., Zaffanella, E., \& Hill, P. (2000).
\newblock Enhanced {S}haring {A}nalysis {T}echniques: {A} {C}omprehensive
  {E}valuation.
\newblock {\em Pages  103--114 of:} {\em Proceedings of {P}rinciples and
  {P}ractice of {D}eclarative {P}rogramming}.
\newblock ACM Press.
\newblock Long version available at http://www.comp.leeds.ac.uk/hill.

\bibitem[\protect\citename{Codish {\em et~al.}\relax, }1991]{CDY91}
Codish, M., Dams, D., \& Yardeni, E. (1991).
\newblock Derivation and {S}afety of an {A}bstract {U}nification {A}lgorithm
  for {G}roundness and {A}liasing {A}nalysis.
\newblock {\em Pages  79--93 of:} {\em Proceedings of the {I}nternational
  {C}onference on {L}ogic {P}rogramming}.
\newblock MIT Press.

\bibitem[\protect\citename{Codish {\em et~al.}\relax, }1999]{CSS99}
Codish, M., S{\o}ndergaard, H., \& Stuckey, P. (1999).
\newblock Sharing and {G}roundness {D}ependencies in {L}ogic {P}rograms.
\newblock {\em {T}ransactions on {P}rogramming {L}anguages and {S}ystems}, {\bf
  21}(5), 948--976.

\bibitem[\protect\citename{Fil\'{e}, }1994]{F94}
Fil\'{e}, G. (1994).
\newblock {\em Sharing $\times$ {F}ree: {S}imple and {C}orrect}.
\newblock Tech. rept.~15. Dipartimento di Matematica, Universit\`{a} Degli
  Studi di Padova.

\bibitem[\protect\citename{Hill {\em et~al.}\relax, }2002]{HBZ02}
Hill, P., Bagnara, R., \& Zaffanella, E. (2002).
\newblock Soundness, {I}dempotence and {C}ommutativity of {S}et-{S}haring.
\newblock {\em Theory and {P}ractice of {L}ogic {P}rogramming}, {\bf 2}(2),
  155--201.

\bibitem[\protect\citename{King, }2000]{K00}
King, A. (2000).
\newblock Pair-{S}haring over {R}ational {T}rees.
\newblock {\em Journal of {L}ogic {P}rogramming}, {\bf 46}(1--2), 139--155.

\bibitem[\protect\citename{Langen, }1991]{L91}
Langen, A. (1991).
\newblock {\em Advanced {T}echniques for {A}pproximating {V}ariable {A}liasing
  in {L}ogic {P}rograms}.
\newblock Ph.D. thesis, University of Southern California, Los Angeles.

\bibitem[\protect\citename{Lassez {\em et~al.}\relax, }1988]{LMM88}
Lassez, J-L., Maher, M., \& Marriott, K. (1988).
\newblock Unification {R}evisited.
\newblock {\em Pages  587--625 of:} {\em Foundations of {D}eductive {D}atabases
  and {L}ogic {P}rogramming}.
\newblock Morgan Kaufmann.

\bibitem[\protect\citename{Muthukumar \& Hermenegildo, }1992]{MH92}
Muthukumar, K., \& Hermenegildo, M. (1992).
\newblock Compile-time {D}erivation of {V}ariable {D}ependency using {A}bstract
  {I}nterpretation.
\newblock {\em Journal of {L}ogic {P}rogramming}, {\bf 13}(2\&3), 315--347.

\end{thebibliography}

\end{document}